\documentclass[11pt,letterpaper]{article}
\usepackage[T1]{fontenc}
\usepackage[latin9]{inputenc}
\usepackage{textcomp}
\usepackage{amsmath}
\usepackage{amssymb}
\usepackage{esint}

\makeatletter

\pdfpageheight\paperheight
\pdfpagewidth\paperwidth

\usepackage{jheppub}

\usepackage{etoolbox}
    
   \patchcmd{\maketitle}{\@fpheader}{}{}{}

\usepackage{amsfonts}
\usepackage{xcolor}

\title{Gravity coupled to a scalar field from a Chern-Simons action: describing rotating hairy black holes and solitons with gauge fields}

\author[a]{Marcela C\'ardenas,}
\author[b]{Oscar Fuentealba,}
\author[c,d]{Cristi\'an Mart\'{\i}nez}
\author[c,d]{and Ricardo Troncoso}

\affiliation[a]{Departamento de F\'{i}sica, Universidad de Santiago de Chile, Avenida V\'{i}ctor Jara 3493, Santiago, Chile}
\affiliation[b]{Universit\'e Libre de Bruxelles and International Solvay Institutes, ULB-Campus Plaine CP231, B-1050 Brussels, Belgium}
\affiliation[c]{Centro de Estudios Cient\'{\i}ficos (CECs),  Av. Arturo Prat 514, Valdivia, Chile}
\affiliation[d]{Facultad de Ingenier\'{i}a, Arquitectura y Dise\~{n}o, Universidad San Sebasti\'an, sede Valdivia, General Lagos 1163, Valdivia 5110693, Chile}

\emailAdd{marcela.cardenas@usach.cl}
\emailAdd{oscar.fuentealba@ulb.be}
\emailAdd{cristian.martinez@uss.cl}
\emailAdd{ricardo.troncoso@uss.cl}

\preprint{CECS-PHY-15/04}

\abstract{Einstein gravity minimally coupled to a scalar field with a two-parameter Higgs-like self-interaction in three spacetime dimensions is recast in terms of a Chern-Simons form for the algebra $g^{+}\oplus g^{-}$ where, depending on the sign of the self-interaction couplings, $g^{\pm}$ can be $so(2,2)$, $so(3,1)$ or $iso(2,1)$. The field equations can then be expressed through the field strength of non-flat composite gauge fields, and conserved charges are readily obtained from boundary terms in the action that agree with those of standard Chern-Simons theory for pure gravity, but with non-flat connections. Regularity of the fields then amounts to requiring the holonomy of the connections along contractible cycles to be trivial. These conditions are automatically fulfilled for the scalar soliton and allow to recover the Hawking temperature and chemical potential in the case of the rotating hairy black holes presented here, whose entropy can also be obtained by the same formula that holds in the case of a pure Chern-Simons theory. In the conformal (Jordan) frame the theory is described by General Relativity with cosmological constant conformally coupled to a self-interacting scalar field, and its formulation in terms of a Chern-Simons form for suitably composite gauge fields is also briefly addressed.} 

\makeatother

\begin{document}
\maketitle \flushbottom

\newpage{}

\section{Introduction}

The formulation of three-dimensional General Relativity as a Chern-Simons
theory \cite{Achucarro:1986uwr,Witten:1988hc} has allowed exploiting
time-honored tools available for gauge fields in order to span a wealth
of very interesting achievements on classical and quantum aspects
of gravitation (for a non-exhaustive list of references, see e.g.
\cite{Carlip:1995zj,Coussaert:1995zp,Maloney:2007ud,Kraus:2006wn,Cotler:2018zff,Perez:2016vqo,Gonzalez:2018jgp,Melnikov:2018fhb,Fuentealba:2017omf,Afshar:2016wfy,Cardenas:2021vwo,Barnich:2013yka,Grumiller:2019tyl,Ojeda:2019xih}).
Many of these results also extend to supergravity \cite{Achucarro:1989gm,Howe:1995zm,Banados:1996hi,Henneaux:1999ib,Giacomini:2006dr,Barnich:2015sca,Fuentealba:2017fck,Caroca:2018obf,Caroca:2019dds,Banerjee:2019lrv,Banerjee:2021uxl,Barnich:2014cwa}
as well as for gravitation coupled to higher-spin fields \cite{Blencowe:1988gj,Bergshoeff:1989ns,Henneaux:2010xg,Campoleoni:2010zq,Gutperle:2011kf,Ammon:2011nk,Castro:2011fm,Henneaux:2012ny,Perez:2012cf,Campoleoni:2012hp,Perez:2013xi,Henneaux:2013dra,Bunster:2014mua,Zinoviev:2014sza,Fuentealba:2015jma,Fuentealba:2015wza,Henneaux:2015tar,Grumiller:2016kcp},
since both can be described in terms of a Chern-Simons theory for
suitable gauge groups. Ultra and non-relativistic versions thereof
have also been developed in \cite{Aviles:2019xed,Ravera:2019ize,Ali:2019jjp,Concha:2020eam,Concha:2021llq,Caroca:2022byi,Grumiller:2017sjh}.

The case of General Relativity minimally coupled to a real scalar
field appears to be far from that kind of description. Nevertheless,
here we show that in the case of a precise two-parameter Higgs-like
self-interaction potential given by 
\begin{equation}
V\left(\phi\right)=\frac{\Lambda}{8}\left(\cosh^{6}\phi+\nu\sinh^{6}\phi\right)\,,\label{eq:PotentialV}
\end{equation}
the theory can be equivalently formulated in terms of a Chern-Simons
form that depends on composite gauge fields, so that the field equations
no longer imply the vanishing of the field strengths. Thus, the theory
and its configuration space can be equivalently described in terms
of non-flat connections, so that many of the gauge theory tools open
up to analyze their properties.

The self-interaction potential \eqref{eq:PotentialV} enjoys some
remarkable properties. Indeed, the first analytic example of a black
hole with a minimally coupled scalar field that circumvents the no-hair
conjecture was precisely found for $V\left(\phi\right)$ in \eqref{eq:PotentialV}
in the range $\Lambda<0$ and $\nu\geq-1$ \cite{Henneaux:2002wm}.
Furthermore, the potential in \eqref{eq:PotentialV} falls within
the class analyzed in \cite{Henneaux:2002wm}, for which the scalar
field acquires a slow fall-off at infinity, so that the canonical
generators of the asymptotic symmetries acquire an explicit contribution
from the scalar field. Thus, it was shown that the Brown-Henneaux
boundary conditions \cite{Brown:1986nw} for gravity with a localized
distribution of matter can be consistently relaxed, so that the asymptotic
symmetries are still given by the conformal group in two dimensions
with the same central extension. As it was also pointed out in \cite{Henneaux:2002wm},
for the self-interaction potential \eqref{eq:PotentialV}, once the
theory is expressed in the conformal (Jordan) frame, the matter piece
of the action becomes conformally invariant, and the action reduces
to General Relativity with cosmological constant $\Lambda$ with a
conformally coupled scalar field, whose self-interaction coupling
is determined by $\nu$ (see section \ref{Ending-remarks}).

The plan of the paper is as follows. In the next section, we show
that the action 
\begin{equation}
I\left[\phi,g_{\mu\nu}\right]=\frac{8}{\kappa}\int d^{3}x\sqrt{-g}\left(\frac{R}{16}-\frac{1}{2}g^{\mu\nu}\partial_{\mu}\phi\partial_{\nu}\phi-V\left(\phi\right)\right)\,,\label{actionmin}
\end{equation}
with the self-interaction potential given by \eqref{eq:PotentialV},
whose field equations read 
\begin{eqnarray}
R_{\mu\nu}-\frac{1}{2}g_{\mu\nu}R & = & 8T_{\mu\nu}\,,\label{eq:Einstein}\\
\Box\phi-\frac{dV\left(\phi\right)}{d\phi} & = & 0\,,\label{eq:Klein-Gordon}
\end{eqnarray}
with 
\begin{equation}
T_{\text{\ensuremath{\mu\nu}}}=\partial_{\mu}\phi\partial_{\nu}\phi-\frac{1}{2}g_{\mu\nu}g^{\alpha\beta}\partial_{\alpha}\phi\partial_{\beta}\phi-g_{\text{\ensuremath{\mu\nu}}}V\left(\phi\right)\,,\label{eq:TensorEM}
\end{equation}
can be expressed in terms of a Chern-Simons form with certain suitable
composite gauge fields. In section \ref{section3}, the Hamiltonian
boundary terms required to obtain the conserved charges in terms of
the connections are worked out, including the hairy black hole entropy
formula in terms of gauge fields. The rotating extension of the hairy
black hole in \cite{Henneaux:2002wm} and its global charges are discussed
in section \ref{section4}. Regularity of hairy black holes and the
scalar soliton in terms of trivial holonomies is addressed in section
\ref{section5}, including the hairy black hole thermodynamics and
the corresponding Cardy formula for its entropy that depends on the
global charges of the scalar soliton. Finally, section \ref{Ending-remarks}
is devoted to some ending remarks as well as a brief discussion of
the formulation of the theory in the conformal frame in terms of a
Chern-Simons form for composite gauge fields.

\section{Action from a Chern-Simons form with composite gauge fields\label{section2}}

Here we show that the action \eqref{actionmin}, up to a boundary
term, can be recast using a Chern-Simons form. The gauge fields are
defined in terms of the direct sum of the algebras $g^{+}$ and $g^{-}$,
where $g^{\pm}$ can be the three-dimensional (anti-)de Sitter or
Poincaré algebras, depending on the signs of $\Lambda^{+}=\Lambda$
and $\Lambda^{-}=-\nu\Lambda$.

The theory can be formulated through the following composite gauge
fields 
\begin{eqnarray}
A^{+} & = & \cosh^{2}\left(\phi\right)e^{a}P_{a}^{+}+\omega_{+}^{a}J_{a}^{+}\,,\label{eq:Ap}\\
A^{-} & = & \sinh^{2}\left(\phi\right)e^{a}P_{a}^{-}+\omega_{-}^{a}J_{a}^{-}\,,\label{eq:Am}
\end{eqnarray}
that depend on the scalar field $\phi$, the dreibein $e^{a}$ and
additional 1-forms $\omega_{\pm}^{a}$. The generators $J_{a}^{\pm}$
and $P_{a}^{^{\pm}}$ fulfill the following algebra 
\begin{equation}
\left[J_{a}^{\pm},J_{b}^{\pm}\right]=\epsilon_{abc}J_{\pm}^{c}\quad,\quad\left[J_{a}^{\pm},P_{b}^{\pm}\right]=\epsilon_{abc}P_{\pm}^{c}\quad,\quad\left[P_{a}^{\pm},P_{b}^{\pm}\right]=-\Lambda^{\pm}\epsilon_{abc}J_{\pm}^{c}\,,\label{eq:algebras}
\end{equation}
so that each copy corresponds to $so(3,1)$ or $so(2,2)$ for positive
or negative signs of $\Lambda^{\pm}$, respectively, or $iso(2,1)$
when $\Lambda^{+}$ or $\Lambda^{-}$ vanishes. The field strengths
associated to the gauge fields \eqref{eq:Ap} and \eqref{eq:Am} are
given by 
\begin{equation}
F^{\pm}=dA^{\pm}+(A^{\pm})^{2}=\mathcal{T}_{\text{\ensuremath{\pm}}}^{a}P_{a}^{\pm}+\mathcal{R}_{\pm}^{a}J_{a}^{\pm}\,,\label{eq:Fpm}
\end{equation}
where the components along the $so(2,1)$ generators $J_{a}^{\pm}$
read 
\begin{eqnarray}
\mathcal{R}_{+}^{a} & = & R_{+}^{a}-\frac{1}{2}\Lambda^{+}\cosh^{4}\left(\phi\right)\epsilon^{abc}e_{b}e_{c}\,,\\
\mathcal{R}_{-}^{a} & = & R_{-}^{a}-\frac{1}{2}\Lambda^{-}\sinh^{4}\left(\phi\right)\epsilon^{abc}e_{b}e_{c}\,,
\end{eqnarray}
with $R_{\pm}^{a}=d\omega_{\text{\ensuremath{\pm}}}^{a}+\frac{1}{2}\epsilon^{abc}\omega_{b}^{\pm}\omega_{c}^{\pm}$,
while those along the remaining generators $P_{a}^{\pm}$ are 
\begin{eqnarray}
\mathcal{T}_{+}^{a} & = & \cosh^{2}\left(\phi\right)\left(T_{+}^{a}+2\tanh\left(\phi\right)d\phi\,e^{a}\right)\,,\\
\mathcal{T}_{-}^{a} & = & \sinh^{2}\left(\phi\right)\left(T_{-}^{a}+2[\tanh\left(\phi\right)]^{-1}d\phi\,e^{a}\right)\,,
\end{eqnarray}
with $T_{\text{\ensuremath{\pm}}}^{a}=de^{a}+\epsilon^{abc}\omega_{b}^{\pm}e_{c}$.

We then consider an action principle defined as a combination of two
Chern-Simons forms for the composite gauge fields $A^{\pm}$, which
reads 
\begin{equation}
I\left[\phi,e,\omega^{+},\omega^{-}\right]=\frac{k^{+}}{4\pi}\int\left\langle A^{+}dA^{+}+\frac{2}{3}(A^{+})^{3}\right\rangle +\frac{k^{-}}{4\pi}\int\left\langle A^{-}dA^{-}+\frac{2}{3}(A^{-})^{3}\right\rangle \,,\label{eq:CS-action}
\end{equation}
where the nonvanishing components of the invariant bilinear form are
given by $\left\langle J_{a}^{\pm},P_{b}^{\pm}\right\rangle =\eta_{ab}$,
with $\eta_{ab}$ standing for the Minkowski metric, and the levels
are defined as $k^{\pm}=\pm2\pi/\kappa$.

The field equations are found varying the action \eqref{eq:CS-action}
with respect to the dynamical fields $e^{a}$, $\phi$ and $\omega_{\pm}^{a}$
\begin{eqnarray}
\delta I & = & \frac{1}{2\pi}\intop\left\langle \left(k^{+}F^{+}\frac{\delta A^{+}}{\delta e^{a}}+k^{-}F^{-}\frac{\delta A^{-}}{\delta e^{a}}\right)\delta e^{a}+\left(k^{+}F^{+}\frac{\delta A^{+}}{\delta\phi}+k^{-}F^{-}\frac{\delta A^{-}}{\delta\phi}\right)\delta\phi\right.\\
 &  & +\left.\left(k^{+}F^{+}\frac{\delta A^{+}}{\delta\omega_{+}^{a}}+k^{-}F^{-}\frac{\delta A^{-}}{\delta\omega_{+}^{a}}\right)\delta\omega_{+}^{a}+\left(k^{+}F^{+}\frac{\delta A^{+}}{\delta\omega_{-}^{a}}+k^{-}F^{-}\frac{\delta A^{-}}{\delta\omega_{-}^{a}}\right)\delta\omega_{-}^{a}\right\rangle \,,
\end{eqnarray}
so that they read 
\begin{eqnarray}
 &  & \cosh^{2}\left(\phi\right)\mathcal{R}_{a}^{+}-\sinh^{2}\left(\phi\right)\mathcal{R}_{a}^{-}=0\,,\label{eq:de}\\
 &  & \left(\mathcal{R}_{a}^{+}-\mathcal{R}_{a}^{-}\right)e^{a}=0\,,\label{eq:dphi}\\
 &  & T_{\pm}^{a}+2[\tanh\left(\phi\right)]^{\pm1}d\phi\,e^{a}=0\,,\label{eq:domegapm}
\end{eqnarray}
respectively.

Note that the last field equations \eqref{eq:domegapm} are algebraic
for $\omega_{\pm}^{a}$, which can be solved as 
\begin{equation}
\omega_{\pm}^{a}\left(e,\phi\right)=\omega^{a}\left(e\right)-2[\tanh\left(\phi\right)]^{\pm1}*\left(e^{a}d\phi\right)\,,\label{eq:wpm}
\end{equation}
where $\omega^{a}$ is the torsionless Levi-Civita spin connection
associated to $e^{a}$, and $*$ stands for the Hodge dual, so that
$*\left(e^{a}d\phi\right)=\epsilon^{abc}\partial_{\nu}\phi e_{b}^{\nu}e_{c}$.

A second order action $I(e^{a},\phi)$ is obtained by replacing \eqref{eq:wpm}
in the Chern-Simons action \eqref{eq:CS-action}, which reduces to
\eqref{actionmin} up to a boundary term, where the self-interaction
potential is precisely given by \eqref{eq:PotentialV}. Analogously,
substituting equation \eqref{eq:wpm} in \eqref{eq:de} and \eqref{eq:dphi},
they reduce respectively to the Einstein and scalar field equations
\eqref{eq:Einstein} and \eqref{eq:Klein-Gordon}.

\section{Global charges and entropy in terms of gauge fields\label{section3}}

In order to deal with conserved charges as well as for the hairy black
hole entropy, it is useful to express the action \eqref{eq:CS-action}
in Hamiltonian form. Supplementing the action with a boundary term
$B$, that is required to ensure a well-defined variational principle
\cite{Regge:1974zd}, yields to 
\begin{equation}
I=\frac{k^{+}}{4\pi}\int dtd^{2}x\epsilon^{ij}\left\langle \dot{A}_{i}^{+}A_{j}^{+}+A_{t}^{+}F_{ij}^{+}\right\rangle +\frac{k^{-}}{4\pi}\int dtd^{2}x\epsilon^{ij}\left\langle \dot{A}_{i}^{-}A_{j}^{-}+A_{t}^{-}F_{ij}^{-}\right\rangle +\mathcal{B}\,,
\end{equation}
where $F_{ij}^{\pm}$ are the spatial components of the field strengths
$F^{\pm}$ in \eqref{eq:Fpm}. Considering that the variations of
the gauge fields depend on the scalar field and the dreiben, and making
use of \eqref{eq:wpm}, the variation of the action reads 
\begin{eqnarray*}
\delta I & = & \frac{1}{2\kappa}\int dtd^{2}x\epsilon^{ij}\left\langle \left[\left(-2F_{tj}^{+}\frac{\delta A_{i}^{+}}{\delta e^{a}}+F_{ij}^{+}\frac{\delta A_{t}^{+}}{\delta e^{a}}\right)-\left(-2F_{tj}^{-}\frac{\delta A_{i}^{-}}{\delta e^{a}}+F_{ij}^{-}\frac{\delta A_{t}^{-}}{\delta e^{a}}\right)\right]\delta e^{a}\right.\\
 &  & +\left.\left[\left(-2F_{tj}^{+}\frac{\delta A_{i}^{+}}{\delta\phi}+F_{ij}^{+}\frac{\delta A_{t}^{+}}{\delta\phi}\right)-\left(-2F_{tj}^{-}\frac{\delta A_{i}^{-}}{\delta\phi}+F_{ij}^{-}\frac{\delta A_{t}^{-}}{\delta\phi}\right)\right]\delta\phi\right\rangle \\
 &  & +\frac{1}{\kappa}\int dtdS_{i}\epsilon^{ij}\left\langle \left(A_{t}^{+}\frac{\delta A_{j}^{+}}{\delta e^{a}}-A_{t}^{-}\frac{\delta A_{j}^{-}}{\delta e^{a}}\right)\delta e^{a}+\left(A_{t}^{+}\frac{\delta A_{j}^{+}}{\delta\phi}-A_{t}^{-}\frac{\delta A_{j}^{-}}{\delta\phi}\right)\delta\phi\right\rangle +\delta\mathcal{B}\,,
\end{eqnarray*}
where the bulk terms are proportional to the field equations \eqref{eq:de}
and \eqref{eq:dphi}.

The variation of the boundary term is then given by 
\begin{eqnarray}
\delta\mathcal{B} & = & -\frac{1}{\kappa}\int dtdS_{i}\epsilon^{ij}\left\langle \left(A_{t}^{+}\frac{\delta A_{j}^{+}}{\delta e^{a}}-A_{t}^{-}\frac{\delta A_{j}^{-}}{\delta e^{a}}\right)\delta e^{a}+\left(A_{t}^{+}\frac{\delta A_{j}^{+}}{\delta\phi}-A_{t}^{-}\frac{\delta A_{j}^{-}}{\delta\phi}\right)\delta\phi\right\rangle \nonumber \\
 & = & -\frac{1}{\kappa}\int dtd\theta\left\langle A_{t}^{+}\delta A_{\theta}^{+}-A_{t}^{-}\delta A_{\theta}^{-}\right\rangle \,,\label{bt}
\end{eqnarray}
which agrees with that of a pure Chern-Simons theory. Nonetheless,
it should be emphasized that in our case the variations of the gauge
fields are not fully independent. For the class of configurations
that we are interested in, that fulfill the asymptotic conditions
spelled out in \cite{Henneaux:2002wm}, it can be also shown that
the variation of the surface integral in \eqref{bt} integrates as
\begin{equation}
\mathcal{B}=-\frac{1}{2\kappa}\int dtd\theta\left\langle A_{t}^{+}A_{\theta}^{+}-A_{t}^{-}A_{\theta}^{-}\right\rangle \,.\label{eq:B1}
\end{equation}

Intriguingly, if one replaces the explicit form of the composite gauge
fields $A^{\pm}$ in \eqref{eq:Ap} and \eqref{eq:Am}, by virtue
of \eqref{eq:wpm}, the boundary term \eqref{eq:B1} completely gets
rid of the scalar field and precisely reduces to that of pure General
Relativity in the standard Chern-Simons formulation, given by 
\begin{equation}
\mathcal{B}=-\frac{1}{2\kappa}\int dtd\theta\left\langle A_{t}A_{\theta}\right\rangle \,,\label{eq:B2}
\end{equation}
where the connection $A$ is defined by setting $\phi=0$ in the definition
of $A^{+}$, i.e., 
\begin{equation}
A=\left.A^{+}\right|{}_{\phi=0}=e^{a}P_{a}^{+}+\omega^{a}J_{a}^{+}\,,\label{eq:Apure}
\end{equation}
stands for the gauge field of pure General Relativity \cite{Achucarro:1986uwr,Witten:1988hc}.

Analogously, the black hole entropy formula that applies for a generic
Chern-Simons theory in \cite{Bunster:2014mua} (see also \cite{deBoer:2013gz}),
reduces to that of General Relativity 
\begin{align}
S & =\frac{1}{\kappa}\oint d\theta\left\langle A_{\tau}^{+}A_{\theta}^{+}-A_{\tau}^{-}A_{\theta}^{-}\right\rangle \Big|_{r_{+}}\nonumber \\
 & =\frac{1}{\kappa}\oint d\theta\left\langle A_{\tau}A_{\theta}\right\rangle \Big|_{r_{+}}\,,\label{eq:SCS}
\end{align}
where $\tau=-it$ is the Euclidean time and the event horizon is located
at $r=r_{\text{+}}$.

In sum, the boundary term \eqref{eq:B2} and the entropy \eqref{eq:SCS}
are found to exclusively depend on the pure gravity gauge field $A$
in \eqref{eq:Apure}, which noteworthy encodes all of the relevant
information, without requiring the contribution of the scalar field
in an explicit way. Indeed, it has to be stressed that the pure gravity
gauge field \eqref{eq:Apure} in our case is generically no longer
flat.

\section{Rotating hairy black hole\label{section4}}

The first analytic example of a black hole solution endowed with minimally
coupled scalar hair was found in \cite{Henneaux:2002wm}, precisely
for the self-interaction under discussion \eqref{eq:PotentialV} in
the case of $\Lambda<0$ and $\nu\geq-1$. Some of its properties
have been further analyzed in \cite{Barnich:2002pi,Clement:2003sr,Gegenberg:2003jr,Park:2004yk,Banados:2005hm,Myung:2008ze,Correa:2010hf,Lashkari:2010ak,Correa:2011dt,Hyun:2012bc,Aparicio:2012yq,Xu:2014qaa,Xu:2014xqa,Ahn:2015uza,Aviles:2018vnf}
following different approaches, and the rotating extension in the
case of $\nu=0$ was presented in \cite{Correa:2012rc}.

In this section we extend the rotating hairy black hole solution to
the full allowed range of the self-interaction coupling ($\nu>-1$).
It can be readily obtained from the static one by performing a Lorentz
boost parameterized by $\omega$, with $\omega^{2}<1$, in the $t-\theta$
cylinder\footnote{The case of $\nu=-1$ is excluded since, as pointed out in \cite{Henneaux:2002wm},
the static metric is invariant under this kind of boosts. This configuration
shares the causal structure with the massless BTZ black hole, but
the null curvature singularity coincides with the ``horizon'' (NUT)
at $r=0$, possessing vanishing mass, angular momentum, temperature
and entropy, regardless the value of the integration constant. }. The line element then reads 
\begin{equation}
ds^{2}=-N_{\infty}^{2}N\left(r\right){}^{2}dt^{2}+\frac{dr^{2}}{G\left(r\right){}^{2}}+R\left(r\right)^{2}\left(d\theta+N^{\theta}\left(r\right)dt\right)^{2}\,,\label{eq:Metric}
\end{equation}
with 
\begin{eqnarray}
N\left(r\right)^{2} & = & \frac{r^{2}g\left(r\right){}^{2}}{R\left(r\right){}^{2}}\,,\\
G\left(r\right)^{2} & = & \left(\frac{H\left(r\right)+2B}{H\left(r\right)+B}\right)^{2}F\left(r\right)^{2}\,,\\
R\left(r\right)^{2} & = & \frac{r^{2}-g\left(r\right)^{2}\omega^{2}\ell^{2}}{1-\omega^{2}}\,,\\
N^{\theta}\left(r\right) & = & N_{\infty}^{\theta}-\frac{\omega}{\ell}\left(\frac{r^{2}-g\left(r\right)^{2}\ell^{2}}{r^{2}-g\left(r\right)^{2}\omega^{2}\ell^{2}}\right)N_{\infty}\,,
\end{eqnarray}
where 
\begin{equation}
g\left(r\right)^{2}=\left(\frac{H\left(r\right)}{H\left(r\right)+B}\right)^{2}F\left(r\right)^{2}\quad,\quad F\left(r\right)^{2}=\frac{H\left(r\right)^{2}}{\ell^{2}}-\left(1+\nu\right)\left(\frac{3B^{2}}{\ell^{2}}+\frac{2B^{3}}{\ell^{2}H\left(r\right)}\right)\,,
\end{equation}
and 
\begin{equation}
H\left(r\right)=\frac{1}{2}\left(r+\sqrt{r^{2}+4Br}\right)\,.
\end{equation}
Here the cosmological constant is written in terms of the AdS radius
as $\Lambda=-\ell^{-2}$, while $N_{\infty}$ and $N_{\infty}^{\theta}$
stand for integration constants that turn out to be related to the
Hawking temperature and the chemical potential, respectively (see
section \ref{section5}). The hairy black hole is dressed with a scalar
field given by 
\begin{equation}
\phi\left(r\right)=\mbox{arctanh}\sqrt{\frac{B}{H(r)+B}}\,,\label{eq:Scalar}
\end{equation}
being real provided that $B>0$. In this case, the scalar field is
regular everywhere except at the origin, since it diverges as $\phi_{r\rightarrow0}=-\frac{1}{4}\log(r)+\cdots$,
sourcing a null curvature singularity, as it is reflected in the behavior
of the Ricci scalar near $r=0$, which reads 
\begin{equation}
R_{r\rightarrow0}=-\frac{4(\nu+1)B^{5/2}}{\ell^{2}r^{5/2}}+O(r^{-3/2})\,.
\end{equation}
Singularities in the spacetime metric and scalar field at the origin
are cloaked by the event horizon located at $r=r_{+}=B\Theta_{\nu}$,
with $\Theta_{\nu}$ given by 
\begin{equation}
\Theta_{\nu}=2\left(z\bar{z}\right)^{2/3}\frac{z^{2/3}-\bar{z}^{2/3}}{z-\bar{z}}\,,
\end{equation}
and $z=1+i\sqrt{\nu}$.

It is worth highlighting that, unlike the case of the rotating solution
in vacuum, the inner region of the rotating hairy black hole does
not possess neither a region with closed timelike curves to be excised
nor an inner (Cauchy) horizon. Indeed, the analogue of the latter
actually corresponds to the null singularity at the origin. Moreover,
the asymptotic behavior of the rotating hairy black hole does not
fulfill the Brown-Henneaux boundary conditions \cite{Brown:1986nw}
because the scalar field \eqref{eq:Scalar} has a slow fall-off at
infinity, generating a strong backreaction on the metric in the asymptotic
region. Instead, the solution fits within the relaxed asymptotic behavior
described in \cite{Henneaux:2002wm}.

In order to describe the rotating hairy black hole in terms of gauge
fields, we choose the local frame so that the dreibein reads 
\begin{equation}
e^{0}=N_{\infty}N\left(r\right)dt\quad,\quad e^{1}=\frac{dr}{G\left(r\right)}\quad,\quad e^{2}=R\left(r\right)\left(d\theta+N^{\theta}\left(r\right)dt\right)\,,\label{eq:Dreibein}
\end{equation}
and hence, the components of the spin connection $\omega^{a}$ are
given by 
\begin{eqnarray}
\omega^{0} & = & G(r)R(r)\mbox{\ensuremath{'}}d\theta+\frac{G(r)}{2}\left(2N^{\theta}(r)R(r)\mbox{\ensuremath{'}}+R(r)N^{\theta}(r)\mbox{\ensuremath{'}}\right)dt\,,\\
\omega^{1} & = & \frac{R(r)N^{\theta}(r)\mbox{\ensuremath{'}}}{2N_{\infty}N\left(r\right)}dr\,,\\
\omega^{2} & = & -\frac{G(r)R(r)^{2}N^{\theta}(r)\mbox{\ensuremath{'}}}{2N_{\infty}N\left(r\right)}d\theta-\frac{G(r)}{2}\left(\frac{R(r)^{2}N^{\theta}(r)N^{\theta}(r)\mbox{\ensuremath{'}}}{N_{\infty}N\left(r\right)}-2N_{\infty}N\left(r\right)\mbox{\ensuremath{'}}\right)dt\,.
\end{eqnarray}
The gauge fields $A^{+}$ and $A^{-}$ in \eqref{eq:Ap}, \eqref{eq:Am}
become then fully specified by virtue of $\omega_{\pm}^{a}$ given
by \eqref{eq:wpm}. Nevertheless, as pointed out in the previous section,
the pure gravity connection in \eqref{eq:Apure} is enough in order
to evaluate the boundary term in \eqref{eq:B2}, which for the rotating
hairy black hole reduces to 
\begin{equation}
\mathcal{B}=-(t_{2}-t_{1})\left[N_{\infty}\left(\frac{3\pi\left(1+\nu\right)B^{2}\left(1+\omega^{2}\right)}{\kappa\ell^{2}\left(1-\omega^{2}\right)}\right)-N_{\infty}^{\theta}\left(\frac{6\pi\left(1+\nu\right)B^{2}\omega}{\kappa\ell\left(1-\omega^{2}\right)}\right)\right].\label{btres}
\end{equation}
Noteworthy, the result is independent of the radial coordinate $r$,
and hence it can be computed even at finite proper distance ($r=r_{0}$)
without the need of taking the limit $r\rightarrow\infty$. The mass
$M$ and angular momentum $J$ can then be recognized from the boundary
term \eqref{btres} in the following way \cite{Regge:1974zd} 
\begin{equation}
\mathcal{B}=(t_{2}-t_{1})\left(-N_{\infty}M+N_{\infty}^{\theta}J\right)\,,\label{btgen}
\end{equation}
so that 
\begin{equation}
M=\frac{3\pi\left(1+\nu\right)B^{2}\left(1+\omega^{2}\right)}{\kappa\ell^{2}\left(1-\omega^{2}\right)}\qquad\mbox{and}\qquad J=\frac{6\pi\left(1+\nu\right)B^{2}\omega}{\kappa\ell\left(1-\omega^{2}\right)}\,.\label{MyJ}
\end{equation}
The mass and angular momentum in \eqref{MyJ} reduce to the result
in \cite{Henneaux:2002wm} for the static case ($\omega=0$), as well
as to that in \cite{Correa:2012rc} for $\nu=0$.

Note that the following bound is fulfilled 
\begin{equation}
\frac{M}{|J|/\ell}=\frac{1+\omega^{2}}{2|\omega|}\geq1\,,
\end{equation}
for $\omega^{2}\leq1$, being saturated at the extremal case ($\omega^{2}=1$).
The extremal case can be attained from \eqref{eq:Metric} and \eqref{eq:Scalar}
by first rescaling the integration constant $B$ according to $b=\frac{B}{\sqrt{1-\omega^{2}}}$,
and then taking the limit $\omega\rightarrow\pm1$, so that the scalar
field vanishes and the line element reduces to that of the extremal
rotating BTZ black hole, whose (degenerate) horizon locates at $r_{+}^{2}=3b^{2}(1+\nu)$.
It is worth noting that static and stationary extremal cases, respectively
for $M=J=0$, and $M=|J|/\ell$, do not admit scalar hair.

\section{Regularity and thermodynamics through non-flat connections \label{section5}}

Here we carry out the thermodynamic analysis of the rotating hairy
black hole relying on its description in terms of gauge fields in
the Euclidean approach, where Euclidean time $\tau=-it$ is normalized
with period equals to 1. The Euclidean hairy black hole in the non-extremal
case possesses the topology of a solid torus, $\mathbb{R}^{2}\times S^{1}$,
where $S^{1}$ is the circle parametrized by the angular coordinate
$\theta$, and $\mathbb{R}^{2}$ stands for the $r-\tau$ plane described
in polar coordinates centered at $r=r_{+}$. Since thermal cycles
around the horizon are contractible, regularity of the gauge fields
at the horizon amounts to require trivial holonomies along them, allowing
to fix the Lagrange multipliers at the boundary, given by $N_{\infty}$
and $N_{\infty}^{\theta}$, which correspond to the Hawking temperature
and the chemical potential.

In concrete, the regularity condition for the Euclidean rotating hairy
black hole is obtained from demanding that the corresponding holonomies
$\mathcal{H^{\pm}}$ along the thermal cycle at the event horizon
to be trivial, i.e., 
\begin{equation}
\mathcal{H^{\pm}}=\exp\left[\intop_{0}^{1}A_{\tau}^{\pm}d\tau\right]_{r_{+}}=\exp\left[A_{\tau}^{\pm}\right]_{r_{+}}=I_{c}\,,\label{eq:Holonomy}
\end{equation}
where $I_{c}$ is a suitable element of the center of the gauge group.

One possible way to implement the regularity condition corresponds
to the direct diagonalization of the holonomies $\mathcal{H}^{\pm}$
for an appropriate matrix representation of the full algebra $g^{+}\oplus$
$g^{-}$. A simpler option is the one spelled out in \cite{Matulich:2014hea},
being implemented in two steps:

(i) Finding a group element that allows to gauge away the temporal
component of the dreibein. In our case this condition is already implemented
since the local Lorentz frame has already been chosen, and $e_{\tau}$
vanishes at the horizon provided that $N^{\theta}(r_{+})=0$, so that
the chemical potential $N_{\infty}^{\theta}$ is fixed in terms of
$N_{\infty}$ as 
\begin{equation}
N_{\infty}^{\theta}=\frac{\omega}{\ell}N_{\infty}\,,\label{nphi-1}
\end{equation}
and the time component of the gauge fields reduce to $A_{\tau}^{\pm}(r_{+})=\omega_{\pm\tau}(r_{+})$,
with $\omega_{\pm\tau}^{a}(r_{+})=\omega_{\tau}^{a}(r_{+})$, whose
non-vanishing component is given by 
\begin{equation}
\omega_{\tau}=N_{\infty}G(r_{+})N\mbox{\ensuremath{'}}(r_{+})J_{2}^{+}\,.
\end{equation}
Therefore, as naturally expected from the results in section \ref{section3},
the problem actually reduces to requiring a trivial holonomy for the
pure gravity gauge field \eqref{eq:Apure}, whose time component at
the horizon reads $A_{\tau}=\omega_{\tau}^{a}(r_{\text{+}})J_{a}^{+}$.

(ii) Diagonalizing the remaining components, which in this case reduces
to that of the corresponding $so(2,1)\approx sl(2,\mathbb{R})$ subalgebra.
The basis can be chosen as

\begin{equation}
J_{0}^{+}=\frac{1}{2}\left(L_{-1}+L_{1}\right),\quad J_{1}^{+}=\frac{1}{2}\left(L_{-1}-L_{1}\right),\quad J_{2}^{+}=L_{0}\,,\label{eq:JsLs}
\end{equation}
with 
\begin{equation}
L_{-1}=\left(\begin{array}{cc}
0 & 0\\
1 & 0
\end{array}\right)\quad,\quad L_{0}=\left(\begin{array}{cc}
-\frac{1}{2} & 0\\
0 & \frac{1}{2}
\end{array}\right)\quad,\quad L_{1}=\left(\begin{array}{cc}
0 & -1\\
0 & 0
\end{array}\right)\,,\label{eq:LsRep}
\end{equation}
fulfilling $\left[L_{n},L_{m}\right]=\left(n-m\right)L_{n+m}$. Since
the $sl(2,\mathbb{R})$ generators are given in the fundamental (spinorial)
representation, the suitable element of the center of the group turns
out to be $I_{c}=-\mathbb{I}$.

The diagonalization can be then readily performed, which implies that
the eigenvalues of $\omega_{\tau}$ must be given by $\pm i\pi$,
or equivalently, 
\begin{equation}
tr\left[(\omega_{\tau})^{2}\right]=-2\pi^{2}\,.\label{eq:TraceCond}
\end{equation}
The latter equation allows to fix the form of $N_{\infty}$, and hence,
by virtue of the former condition in \eqref{nphi-1}, the Lagrange
multipliers, related to the Hawking temperature and the chemical potential,
become fixed as 
\begin{equation}
N_{\infty}=\frac{2\pi\Theta_{\nu}\ell^{2}}{3\left(1+\nu\right)B\sqrt{1-\omega^{2}}}\qquad\mbox{and}\qquad N_{\infty}^{\theta}=\frac{2\pi\Theta_{\nu}\omega\ell}{3\left(1+\nu\right)B\sqrt{1-\omega^{2}}}\,.\label{N4}
\end{equation}
Once the Langrange multipliers are fixed as in \eqref{N4}, the pure
gravity gauge field takes the following form at the horizon 
\begin{align}
A= & R(r_{+})d\theta P_{2}^{+}+\left(2\pi dt-\frac{G(r_{+})^{2}R(r_{+})^{2}N\mbox{\ensuremath{'}}(r_{+})N^{\theta}\mbox{\ensuremath{'}}(r_{+})}{4\pi N(r_{+})}d\theta\right)J_{2}^{+}\,,
\end{align}
so that the rotating hairy black hole entropy can be directly obtained
from \eqref{eq:SCS}, being given by 
\begin{equation}
S=\frac{4\pi^{2}B\Theta_{\nu}}{\kappa\sqrt{1-\omega^{2}}}=\frac{4\pi^{2}R(r_{+})}{\kappa}=\frac{A_{\textrm{hor}}}{4G},\label{BHEntropy}
\end{equation}
with $\kappa=8\pi G$, which is in full agreement with the Bekenstein-Hawking
area law for the entropy. It is then simple to verify that the first
law is fulfilled in the grand canonical ensemble, $dS=\beta dM-\beta\Omega dJ,$
with $M$ and $J$ determined by \eqref{MyJ}, so that the relationship
between the Lagrange multipliers with the Hawking temperature and
the angular velocity at the horizon reads 
\begin{equation}
\beta=N_{\infty}\qquad\textrm{and}\qquad\beta\Omega=N_{\infty}^{\theta}\,.\label{potentials}
\end{equation}

\subsection{Soliton mass, regularity and Cardy formula}

An analytic soliton solution endowed with a nontrivial scalar field
for the theory under discussion was found in \cite{Correa:2010hf}.
The line element is given by 
\begin{eqnarray}
ds^{2} & = & \ell^{2}\left(1+\frac{1}{\alpha_{\nu}\left(1+\rho^{2}\right)}\right)^{-2}\times\nonumber \\
 &  & \left[-N_{\infty}^{2}\left[\frac{2\left(1+\rho^{2}\right)}{3c_{\nu}\ell}\right]^{2}dt^{2}+\frac{4d\rho^{2}}{2+\rho^{2}+\frac{c_{\nu}}{1+\rho^{2}}}+\left(\frac{2\rho}{2+c_{\nu}}\right)^{2}\left(2+\rho^{2}+\frac{c_{\nu}}{1+\rho^{2}}\right)d\theta^{2}\right]\,,\nonumber \\
\end{eqnarray}
with $c_{\nu}=2\alpha_{\nu}^{-3}\left(1+\nu\right)$ and $\alpha_{\nu}=\frac{1}{2}\left(\Theta_{\nu}+\sqrt{\Theta_{\nu}^{2}+4\Theta_{\nu}}\right)$,
while the scalar field reads 
\begin{equation}
\phi\left(\rho\right)=\mbox{arctanh}\left(\sqrt{\frac{1}{1+\alpha_{\nu}\left(1+\rho^{2}\right)}}\right)\,.
\end{equation}
The radial coordinate has the range $0\leq\rho<\infty$, and the remaining
coordinates range precisely as for the hairy black hole metric, so
that the solitonic configuration is regular everywhere. As in the
case of the rotating hairy black hole, the fall-off at infinity of
the soliton also fits that of the relaxed boundary conditions in \cite{Henneaux:2002wm}.

Regularity of the soliton at the origin can also be explicitly verified
in terms of its corresponding gauge fields in \eqref{eq:Ap}, \eqref{eq:Am}.
In order to do that, we follow the same lines as in the case of the
Euclidean hairy rotating black hole discussed above, adapted to this
case. Thus, for the solitonic configuration, the suitable condition
turns out to be requiring the holonomy of the gauge fields $A^{\pm}$
around a spatial (angular) cycle that encloses the origin ($\rho=0$)
to be trivial, i.e., 
\begin{equation}
\mathcal{H^{\pm}}=e^{\oint d\theta A_{\theta}^{\pm}\Big|_{\rho=0}}=\mathbb{I}_{c}\,.
\end{equation}
The local frame can be suitably chosen as in the case of the hairy
black hole, so that the angular components of the dreibein $e_{\theta}$
automatically vanish at the origin, without the need of imposing any
condition. Thus, the angular components of the gauge fields now fulfill
$A_{\theta}^{\pm}\Big|_{\rho=0}=\omega_{\pm\theta}\Big|_{\rho=0}$,
with $\omega_{\pm\theta}^{a}\Big|_{\rho=0}=\omega_{\theta}^{a}\Big|_{\rho=0}$,
so that the problem again reduces to just requiring the trivial holonomy
for the pure gravity connection in \eqref{eq:Apure}, whose angular
component at the origin is given by $A_{\theta}\Big|_{\rho=0}=\omega_{\theta}^{a}J_{a}^{+}\Big|_{\rho=0}$.
The remaining components can then be diagonalized by choosing the
basis and the representation as in \eqref{eq:JsLs} and \eqref{eq:LsRep},
respectively, so that the holonomy turns out to be trivial provided
that the condition in \eqref{eq:TraceCond} with $\omega_{\tau}\rightarrow\omega_{\theta}$
holds, which automatically does without the need of requiring any
additional condition.

The result is reassuring because the soliton is devoid of integration
constants, and hence, it has no freedom to be adjusted in order to
fulfill the regularity conditions. Besides, as pointed out in \cite{Correa:2010hf},
this last feature of the soliton suggests that it can be naturally
regarded as a ground state of the theory for the sector with non-vanishing
scalar fields. Indeed, it is simple to extend the result in \cite{Correa:2010hf}
to the rotating case, so that the Euclidean hairy black hole turns
out to be diffeomorphic to the Euclidean soliton provided that the
corresponding modular parameters of the torus are related through
S-duality, i.e., 
\[
\tau_{sol}=-\frac{1}{\tau_{hbh}}
\]
with $\tau_{hbh}=\frac{i\beta(1-i\Omega)}{2\pi}$, precisely as it
occurs for the Euclidean BTZ black hole \cite{Banados:1992wn,Banados:1992gq}
and Euclidean AdS$_{3}$ \cite{Carlip:1994gc,Maldacena:1998bw}\footnote{Analogue S-duality relationships between Euclidean three-dimensional
black holes and their corresponding diffeomorphic Euclidean solitons
are also know to hold for General Relativity on AdS$_{3}$ with boundary
conditions of KdV type \cite{Perez:2016vqo}, as well as for asymptotically
AdS \cite{Oliva:2009ip} or asymptotically Lifshitz black holes \cite{Ayon-Beato:2009rgu}
and their corresponding solitons in \cite{Perez:2011qp} and \cite{Gonzalez:2011nz},
respectively in the context of BHT massive gravity \cite{Bergshoeff:2009hq}. }.

Thus, assuming that the soliton is the ground state of the hairy sector
of the theory, the entropy of the rotating hairy black hole can be
seen to be successfully reproduced by the Cardy formula once expressed
in terms of left and right ground state energies $\tilde{\Delta}_{0}^{\pm}$
instead of the central charges \cite{Correa:2010hf} (see also \cite{Correa:2011dt,Correa:2012rc}).
The entropy then reads 
\begin{equation}
S=4\pi\sqrt{-\tilde{\Delta}_{0}^{+}\tilde{\Delta}^{+}}+4\pi\sqrt{-\tilde{\Delta}_{0}^{-}\tilde{\Delta}^{-}}\,,\label{eq:Cardy}
\end{equation}
where 
\begin{equation}
\tilde{\Delta}^{\pm}=\frac{1}{2}\left(M\ell\pm J\right)\,,
\end{equation}
stand for the eigenvalues of the shifted Virasoro operators, $\tilde{L}_{0}^{\pm}=L_{0}^{\pm}-\frac{c^{\pm}}{24}$,
being related to the mass $M$ and angular momentum $J$ of the hairy
black hole in \eqref{MyJ}, while left and right energies of the ground
state are determined by the soliton mass according to $\tilde{\Delta}_{0}^{\pm}=\frac{1}{2}\ell M_{sol}$.

The mass of the soliton can then be readily obtained by plugging its
associated pure gravity gauge field \eqref{eq:Apure} into the boundary
term in \eqref{eq:B2}, which yields to 
\begin{equation}
\mathcal{B}=-\frac{1}{2\kappa}\left(t_{2}-t_{1}\right)\oint d\theta\left\langle A_{t}A_{\theta}\right\rangle =\left(t_{2}-t_{1}\right)N_{\infty}\left(\frac{\pi\alpha_{\nu}^{4}}{3\kappa\left(1+\nu\right)\left(1+\alpha_{\nu}\right)^{2}}\right)\,,
\end{equation}
so that the soliton mass is found to be given by 
\begin{equation}
M_{sol}=-\frac{\pi\alpha_{\nu}^{4}}{3\kappa\left(1+\nu\right)\left(1+\alpha_{\nu}\right)^{2}}\,,\label{eq:Msol}
\end{equation}
in agreement the the result obtained in \cite{Correa:2010hf} through
the canonical approach. Therefore, making use of the precise value
of the soliton mass in \eqref{eq:Msol}, it is straightforward to
verify that the Cardy formula \eqref{eq:Cardy} precisely reproduces
the rotating hairy black hole entropy in \eqref{BHEntropy}.

\section{Conformal frame and ending remarks \label{Ending-remarks}}

As pointed out in the introduction, the action of a minimally coupled
scalar field \eqref{eq:PotentialV} with the self-interaction potential
\eqref{eq:PotentialV} relates to that of a conformally coupled self-interacting
scalar field. Indeed, changing to the conformal (Jordan) frame according
to $\hat{g}_{\mu\nu}=\Omega^{-2}g_{\mu\nu}$, with $\Omega=(1-\varphi^{2}),$
and redefining the scalar field as $\varphi=\tanh\left(\phi\right)$,
the action becomes 
\begin{equation}
I\left[\hat{g}_{\mu\nu},\varphi\right]=\frac{8}{\kappa}\int d^{3}x\sqrt{-\hat{g}}\left(\frac{\hat{R}-2\Lambda}{16}-\frac{1}{2}\hat{g}^{\mu\nu}\partial_{\mu}\varphi\partial_{\nu}\varphi-\frac{1}{16}\hat{R}\varphi^{2}-\lambda\varphi^{6}\right)\,,\label{eq:CC-action-1}
\end{equation}
with $\lambda=\frac{\Lambda\nu}{8}$, so that the matter piece turns
out to be conformally invariant. The field equations then read 
\begin{eqnarray}
\hat{G}_{\mu\nu}+\Lambda\hat{g}_{\mu\nu} & = & 8\hat{T}_{\mu\nu}\,,\label{eq:eq of motion einstein}\\
\hat{\Box}\varphi-\frac{1}{8}\hat{R}\varphi-6\lambda\varphi^{5} & = & 0\,,\label{eq:SFeq}
\end{eqnarray}
where the stress-energy tensor 
\begin{equation}
\hat{T}_{\mu\nu}=\partial_{\mu}\varphi\partial_{\nu}\varphi-\frac{1}{2}\hat{g}_{\mu\nu}\partial_{\alpha}\varphi\partial^{\alpha}\varphi-\hat{g}_{\mu\nu}\lambda\varphi^{6}+\frac{1}{8}\left(\hat{g}_{\mu\nu}\hat{\Box}-\hat{\nabla}_{\mu}\hat{\nabla}_{\nu}+\hat{G}_{\mu\nu}\right)\varphi^{2},\label{eq:EM-tensor-1}
\end{equation}
is traceless by virtue of \eqref{eq:SFeq}, implying that the Ricci
scalar is constant, $\hat{R}=-6\ell^{-2}$.

The case of a conformally coupled self-interacting scalar field on
a fixed background metric can also be described in terms of a Chern-Simons
action \cite{Ricardo-Minas}, and here we extend this result to the
case of a back-reacting scalar field with a dynamical metric described
by \eqref{eq:CC-action-1}.

It is then useful to define $\Lambda^{+}=\Lambda$ and $\Lambda^{-}=-8\lambda$,
as well as the appropriate composite gauge fields as 
\begin{eqnarray}
A^{+} & = & \hat{e}^{a}P_{a}^{+}+\omega_{+}^{a}J_{a}^{+}\,,\label{eq:Ap-1}\\
A^{-} & = & \varphi^{2}\hat{e}^{a}P_{a}^{-}+\omega_{-}^{a}J_{a}^{-}\,,\label{eq:Am-1}
\end{eqnarray}
where according to the signs of $\Lambda^{\pm}$, they take values
on the algebras in \eqref{eq:algebras} that may correspond to $so(3,1)$,
$so(2,2)$ or $iso(2,1)$, precisely as in the case of the minimally
coupled scalar field. The field strengths then read 
\begin{equation}
F^{\pm}=dA^{\pm}+(A^{\pm})^{2}=\mathcal{T}_{\text{\ensuremath{\pm}}}^{a}P_{a}^{\pm}+\mathcal{R}_{\pm}^{a}J_{a}^{\pm}\,,\label{eq:Fpm-1}
\end{equation}
whose components are given by 
\begin{eqnarray}
\mathcal{R}_{+}^{a} & = & R_{+}^{a}-\frac{1}{2}\Lambda^{+}\epsilon^{abc}\hat{e}_{b}\hat{e}_{c}\,,\\
\mathcal{R}_{-}^{a} & = & R_{-}^{a}-\frac{1}{2}\Lambda^{-}\varphi^{4}\epsilon^{abc}\hat{e}_{b}\hat{e}_{c}\,,\\
\mathcal{T}_{+}^{a} & = & T_{+}^{a}\,,\\
\mathcal{T}_{-}^{a} & = & \varphi^{2}\left(T_{-}^{a}+2\varphi^{-1}d\varphi\hat{e}^{a}\right)\,,
\end{eqnarray}
where $R_{\pm}^{a}=d\omega_{\text{\ensuremath{\pm}}}^{a}+\frac{1}{2}\epsilon^{abc}\omega_{b}^{\pm}\omega_{c}^{\pm}$,
and $T_{\text{\ensuremath{\pm}}}^{a}=d\hat{e}^{a}+\epsilon^{abc}\omega_{b}^{\pm}\hat{e}_{c}$.

Up to a boundary term, the action \eqref{eq:CC-action-1} can then
be analogously reformulated in terms of a Chern-Simons form as in
\eqref{eq:CS-action}, for the composite gauge fields $A^{\pm}$ in
\eqref{eq:Ap-1}, \eqref{eq:Am-1} so that $I=I\left[\varphi,\hat{e},\omega^{+},\omega^{-}\right]$.
Varying with respect to the dynamical fields one obtains 
\begin{eqnarray}
\frac{\delta I}{\delta\hat{e}^{a}}: &  & \mathcal{R}_{a}^{+}-\varphi^{2}\mathcal{R}_{a}^{-}=0\,,\label{eq:de-1}\\
\frac{\delta I}{\delta\varphi}: &  & \left(\mathcal{R}_{a}^{+}-\mathcal{R}_{a}^{-}\right)\hat{e}^{a}=0\,,\label{eq:dphi-1}\\
\frac{\delta I}{\delta\omega_{+}^{a}}: &  & T_{+}^{a}=0\,,\label{eq:Tplus-conf}\\
\frac{\delta I}{\delta\omega_{-}^{a}}: &  & T_{-}^{a}+2\varphi^{-1}d\varphi\hat{e}^{a}=0\,.\label{eq:domegapm-1}
\end{eqnarray}
Since the last field equations \eqref{eq:Tplus-conf} and \eqref{eq:domegapm-1}
are algebraic for $\omega_{\pm}^{a}$, they are solved as 
\begin{equation}
\omega_{-}^{a}\left(\hat{e},\varphi\right)=\omega^{a}\left(\hat{e}\right)-2\varphi^{-1}*\left(\hat{e}^{a}d\varphi\right)\,.\label{eq:wpm-1}
\end{equation}
and $\omega_{+}^{a}=\omega^{a}\left(\hat{e}\right)$, where $\omega^{a}\left(\hat{e}\right)$
stands for the (torsionless) spin connection associated to $\hat{e}^{a}$.
Making use of them, the remaining field equations \eqref{eq:de-1}
and \eqref{eq:dphi-1} reduce to \eqref{eq:eq of motion einstein}
and \eqref{eq:SFeq}, respectively.

A rotating black hole dressed with a conformally coupled scalar field
can then be readily obtained from the exact solution discussed in
section \ref{section4}, just by transforming from the Einstein to
the conformal frame. In the static case, it reduces to the black holes
discussed in \cite{Henneaux:2002wm} and previously in \cite{Martinez:1996gn}
in the case of $\nu=0$. As in section \ref{section5}, the analysis
of its global charges, regularity and thermodynamics can then also
be performed in terms of the gauge fields $A^{\pm}$ in \eqref{eq:Ap-1},
\eqref{eq:Am-1} by virtue of the boundary terms in \eqref{eq:B1},
\eqref{eq:B2} and the holonomy condition \eqref{eq:Holonomy}. It
is also reassuring to verify that the black hole entropy formula in
the Chern-Simons approach \cite{Bunster:2014mua}, given by \eqref{eq:SCS},
reduces to that in the conformal frame

\[
S=\left[1-\varphi^{2}\big(r_{+}\big)\right]\frac{A_{\textrm{hor}}}{4G}\,,
\]
with $\kappa=8\pi G$, see e.g., \cite{Visser:1993nu,Ashtekar:2003jh},
which can also be reproduced from the Cardy formula \eqref{eq:Cardy}
when the scalar soliton in the conformal frame is assumed to be the
ground state configuration.

As a final remark, it would be interesting to explore whether the
hairy black holes and similar configurations endowed with a non-trivial
scalar field that have been found for a variety of gravity theories,
with different scalar couplings and self-interactions \cite{Hotta:2008xt,Hortacsu:2003we,Kwon:2012zh,Xu:2013nia,Zhao:2013isa,Naji:2014ira,Mazharimousavi:2014vza,Xu:2014uha,Cardenas:2014kaa,Xu:2014uka,Gonzalez:2014pwa,Naji:2014qya,Ayon-Beato:2015jga,Wen:2015xea,Ayon-Beato:2015ada,Fan:2015ykb,Harms:2016pow,Ozcelik:2016scf,Erices:2017izj,Tang:2019jkn,Harms:2017yko,Karakasis:2021lnq,Bueno:2021krl,Karakasis:2021ttn,Arias:2022jax,Desa:2022gtw,Karakasis:2022fep,Bueno:2022ewf},
could also be understood in terms of suitable composite gauge fields.
Nevertheless, the Chern-Simons formulation for the theory discussed
here naively appears to be somewhat rigid. In fact, the form of the
composite gauge fields in \eqref{eq:Ap} and \eqref{eq:Am} can be
extended as $A^{\pm}=f_{\pm}^{2}\left(\phi\right)e^{a}P_{a}^{\pm}+\omega_{\pm}^{a}J_{a}^{\pm}$,
so that consistency of the Chern-Simons formulation with a Riemannian
(torsionless) geometry implies that $f_{+}^{2}-f_{-}^{2}=1$, allows
to obtain General Relativity minimally coupled to a scalar field with
a non-canonical kinetic term and a self-interaction that depend on
$f_{+}(\phi)$. However, the scalar field can be appropriately redefined
so that in terms of the new scalar field one recovers the original
action in \eqref{actionmin} with canonical kinetic term and precisely
the same self-interaction given in \eqref{eq:PotentialV}.

\acknowledgments We would like to thank Hernán González, Marc Henneaux,
Alfredo Pérez, David Tempo and Jorge Zanelli for useful comments and
discussions. This research has been partially supported by ANID FONDECYT
grants N° 11190730, 1201208, 1220862 , 1211226, 1220910, 1221624.
The work of O.F. was partially supported by FNRS-Belgium (conventions
FRFC PDRT.1025.14 and IISN 4.4503.15), as well as by funds from the
Solvay Family.

\appendix

\end{document}